\documentclass[twoside,11pt]{article}

\usepackage{jmlr2e}
\usepackage{enumitem}
\usepackage{amsmath}
\newcommand{\indep}{\rotatebox[origin=c]{90}{$\models$}}

\usepackage{listings}
\usepackage{xcolor}

\definecolor{codegreen}{rgb}{0,0.6,0}
\definecolor{codegray}{rgb}{0.5,0.5,0.5}
\definecolor{codepurple}{rgb}{0.58,0,0.82}
\definecolor{backcolour}{rgb}{0.95,0.95,0.92}

\lstdefinestyle{mystyle}{
    backgroundcolor=\color{backcolour},   
    commentstyle=\color{codegreen},
    keywordstyle=\color{magenta},
    numberstyle=\tiny\color{codegray},
    stringstyle=\color{codepurple},
    basicstyle=\ttfamily\footnotesize,
    breakatwhitespace=false,         
    breaklines=true,                 
    captionpos=b,                    
    keepspaces=true,                 
    numbers=left,                    
    numbersep=5pt,                  
    showspaces=false,                
    showstringspaces=false,
    showtabs=false,                  
    tabsize=2
}
\ShortHeadings{Prespecification of Structure}{Vowels, M.J.}
\firstpageno{1}

\begin{document}

\title{Prespecification of Structure for Optimizing Data Collection and Research Transparency by Leveraging Conditional Independencies (preprint)}

\author{\name Matthew J. Vowels \email matthew.vowels@unil.ch \\
}

\editor{}

\maketitle

\begin{abstract}%   <- trailing '%' for backward compatibility of .sty file
Data collection and research methodology represents a critical part of the research pipeline. On the one hand, it is important that we collect data in a way that maximises the validity of what we are measuring, which may involve the use of long scales with many items. On the other hand, collecting a large number of items across multiple scales results in participant fatigue, and expensive and time consuming data collection. It is therefore important that we use the available resources optimally. In this work, we consider how a consideration for theory and the associated causal/structural model can help us to streamline data collection procedures by not wasting time collecting data for variables which are not causally critical for subsequent analysis. This not only saves time and enables us to redirect resources to attend to other variables which are more important, but also increases research transparency and the reliability of theory testing. 
 In order to achieve this streamlined data collection, we leverage structural models, and Markov conditional independency structures implicit in these models to identify the substructures which are critical for answering a particular research question. In this work, we review the relevant concepts and present a number of didactic examples with the hope that psychologists can use these techniques to streamline their data collection process without invalidating the subsequent analysis. We provide a number of simulation results to demonstrate the limited analytical impact of this streamlining. 
\end{abstract}

% \pagenumbering{gobble} 
\section{Introduction}

The data collection and research methodology stages of a research project pipeline are critical for enabling researchers to answer their research questions or test their hypotheses. Firstly, the scales used must each be adequately comprehensive if they are to measure what they are intended to measure. Using a single item questionnaire to measure depression, for instance, may be ill advised. Secondly, there must be a broad enough range of scales such that it is possible to reflect all the relevant components of a psychological process to facilitate valid, and well-specified statistical inference. For example, exploring relationships between variables often entails the measurement of a number of control variables. Unfortunately, these requirements increase the cost and time required to complete data collection, having an impact on participant fatigue \citep{Lavrakas2008} as well as draining valuable project resources.

In this paper, we argue that the data collection and research project methodology can be optimized through a consideration for the structural model reflecting the underling theory, even if a structural model is not used for the analysis. From an inspection of the structural model, one can ascertain which variables or scales are causally necessary, and which can be omitted from the data collection process. This thereby liberates resources to either improve the quality of the remaining scales (by using scales with a more comprehensive set of items), and/or to reduce participant fatigue by shortening the duration of a questionnaire and using these resources to increase the overall sample size. Indeed, concerns about inadequate statistical power are growing in response to the replication crisis \citep{Sassenberg2019, Baker2020, Correll2020, aarts}, and researchers are thus encouraged to make sure they have sufficient data to estimate the effects of interest.

Furthermore, we argue that the process of reflecting a theory as a structural model helps with transparency, reproducibility, and the meaningfulness of subsequent interpretation. Psychology has been accused of being `not even wrong' \citep{Scheel2022} on the basis that the theories are too vague to be adequately tested. By reflecting our theories in a structural model, we thus improve the clarity and reduce the one-to-many relationship between our theories and our statistical models. Translating our theories to structural models also forces researchers to think carefully about the underlying process, and the concomitant implications for data collection.

In this work we show how four related concepts - Markov conditional independencies, Markov Blankets, projection, and causal identification - can be used to judiciously shrink the size of structural models without impacting downstream analyses and without impacting the congruity of the model with the underlying theory. Indeed, the process is not data-driven and is not the same as seeking model `parsimony' - our approach does not fundamentally change the complexity of the underlying processes reflected by the `full' structural model. Instead, the model is reduced to focus in on the effects we care about using a set of rules which are consistent with the assumptions of the original graph being specified. Thus whilst the complexity of the statistical model reduces, it does so without introducing any additional simplifying assumptions beyond those which already existed in the original theory.

We begin with a discussion of the motivation behind specification of our theories as structural models. Then, we introduce the relevant statistical/structural concepts needed to understand the process for reducing a structural model, assuming the model has been informed by a relevant psychological theory. We the walk through a number of didactic examples, comparing an assumed `real-world' or Data Generating Process (DGP) against the minimal required structural models for estimating a set of causal effects of interest. We also provide the associated multiple linear regression models where a single regression model can be used to provide the same information, and present a set of empirical structural equation modeling results demonstrating the empirical validity of the approach.\footnote{We also provide code for reproducing the simulations, as well as for automatically reducing the graph here: \url{https://anonymous.4open.science/r/minSEM-2338/README.md}}. Our hope is that researchers can use the techniques presented in this work so that they can optimize their data collection and analysis in a way which is more transparent and tailored more specifically to the particular relationships of interest.

\section{Motivation}

Path and Structural Equation Models (referred to jointly as SEMs hereafter) provide us with powerful and popular \citep{Blanca2018} statistical frameworks to unambiguously reflect and test causal theories and relationships \citep{Vowels2021, Rohrer2018, Grosz2020, Wright1921, Wright1923, Pearl2009}. Such frameworks allow for a direct representation of a psychological theory in an intuitive graphical (and therefore visual) representation of domain knowledge regarding the underling process of interest. They can therefore also be used early on in the research pipeline to inform the data collection methodology for a research project, by helping us specify which constructs we need to measure. Furthermore, early specification of a statistical models helps us with preregistration and research transparency \citep{Wagenmakers2012}. Such transparency is increasingly important in the fields of psychology and social science, where attention has been drawn to numerous problems with theory testing, research methodology, and analytical practice \citep{Vowels2021, Flake2020, Scheel2020, Gigerenzer2018, McShane2019, aarts, Marsman2017}. 

We focus on two problems in particular. Firstly, psychological research is frequently \textit{underpowered} \citep{Vankov2014, Maxwell2004, Crutzen2017}, and secondly, the theory and analysis are poorly specified \citep{Scheel2020, Scheel2022, Grosz2020, Rohrer2018, Vowels2021}. The studies are underpowered to the extent that the sample sizes are insufficient to test a target hypothesis. For example, for a minimum assumed \textit{true} effect size of interest, it is generally recommended that enough data are collected to yield a power of 80\%, meaning that there is an 80\% probability that we will find a statistically significant result (at a given threshold such as 0.05) \citep{Gelman2021}. If we know a study is underpowered, a statistically significant result should raise concerns because it will tend to occur by chance (our particular sample has an effect size which is unusually high). Researchers are thus encouraged to ensure that their studies are adequately powered, and have been encouraged to do so for some time \citep{Vankov2014, Sedlmeier1989}. However, depending on the complexity of the theory under test, researchers may need to measure a large number of constructs, each with a large number of items. For example, depending on the format, the IPIP-NEO Big 5 inventory contains between 120-300 items \citep{Goldberg1999, Goldberg2006} and therefore takes considerable time to complete. Besides the associated cost and time required to measure constructs using such comprehensive scales, the participants may also experience fatigue, lowering the quality of the responses \citep{Lavrakas2008}. There thus exists a need to compromise - maximising the quality of a survey such that it measures all that we need, at a sufficient level of quality, for a sufficient number of participants.

The second problem of poor specification has prompted meta-researchers to describe research in psychology as `not even wrong' \citep{Scheel2022}. That is to say, if the theories are too vague to be specified unambiguously, then it is not clear what it is that any particular statistical test is actually testing. Indeed, a single theory may admit multiple statistical models, each of which tests something slightly different but all of which are valid given the theory. The inclusion of different control variables can have a large impact on the resulting parameter estimates, and it is not usually clear how these control variables are chosen or how they relate to the tested theory \citep{Vowels2021, Hullman2022, Cinelli2020}. To this extent, it is important that psychologists consider not only the structure between the primary constructs central to their theory, but also the full data-generating process (DGP) which leads to a set of observations. The theory can then be translated into a structural model which reflects this DGP. The process of deriving a structural model from our theory has been previously discussed by \citet{Rohrer2018} and others \citep{Kline2005, Loehlin2017}, and we do not describe the procedure in this work, but note that the graphical framework (more about this in later sections) makes the process quite intuitive. 

The advantages of reflecting the theory unambiguously in a structural model include reproducibility (it is clear what exactly is being tested) and an increase in the interpretability and validity of the resulting effect sizes. Rather than the effect sizes being arbitrary consequences of \textit{ad hoc} models loosely connected to theory, they reflect specific causal effects within a fully specified structural/causal process. The task of translating our theories may also highlight possible weaknesses in the theory, or call attention to possibly insurmountable difficulties for data collection. For instance, theories which involve dynamic processes that unfold at irregular intervals over time may require very specific, expensive, and challenging data collection procedures \citep{Hilpert2019}. Identifying the specifics of such challenges in advance could save a lot of wasted time and effort.

Unfortunately, the task of identifying all relevant variables will likely implicate a large number of secondary variables (such as demographics and other theoretically related constructs), thus requiring longer surveys. The problems of statistical power, comprehensive scale inventories, and the need to collect a broad range of variables and constructs relevant to our theory thus put a lot of pressure on researchers to find a suitable `Goldilocks' design, and one or multiple methodological facets are likely to be compromised as a consequence. As such, after the specification of the full DGP, we should examine the resulting model to identify possible shortcuts to save in the data collection process. Indeed, and as we will show, even if a variable or construct is relevant to a particular causal process, it may not be required for the subsequent analysis. In order to know this, however, the variable needs to be transparently situated in a causal model for us to understand whether it is essential or not. 

Once the structure of the DGP is fully specified, and as we will describe in detail below, we are able to identify essential substructures which are sufficient for testing our intended hypotheses. The substructures, by definition, exclude certain variables. Thus, if we can identify these substructures in advance of data collection, we may be able to significantly reduce the number of constructs we need to measure. Indeed, in example 2i in Figure \ref{fig:all}) below, we show that it is possible to reduce the number of variables/constructs by two thirds, although this depends on how much of the process we are interested in testing. It goes without saying that any simplification must be done carefully. Indeed, the potential consequences of any resultant model misspecification can be severe, and includes heavily biased parameter estimates which are almost impossible to meaningfully interpret \citep{Vowels2021, Hullman2022}.

We thus advocate that researchers consider the DGP upfront, before the data collection stage. Such
prespecification in the form of a structural model represents a beneficial step in terms of preregistration and transparency, helps researchers distill their theories into testable models, thereby increasing the validity and meaningfulness of downstream statistical inference and results interpretations, and provides us with an opportunity to `prune' the structure to optimize for statistical power during data collection.

 \section{Background}
 In this section, we introduce a number of relevant technical concepts for reducing our structural models. In general, we assume that the model is being specified as a path model, or a Structural Equation Model (SEM), and for convenience we consider these two frameworks as equivalent. This is because both types represent subtypes of the graphical modeling framework  known as Probabilistic Graphical Models \citep{Koller2009, Pearl2009}, and the relevant concepts are adequately reflected in SEMs which are already popular in psychology \citep{Blanca2018}. In order to avoid terminological pedantry we thus assume researchers are using SEMs, but note that the ideas here generalise to other structural frameworks as well (such as Directed Acyclic Graphs and Structural Causal Models).

A number of existing resources exist which discuss the implications of changes in causal structures on statistical estimation. For example, \citet{Vowels2021} discusses the problems that arise due to misspecification of causal models, and briefly notes the potential to focus on specific effects within a causal process; and \citet{Cinelli2020} provides a laconic summary of how to choose control variables such that the choice does not induce bias in our parameter estimation. However, these resources do not discuss the possibility of reducing our SEMs to the most simple model which can still yield unbiased estimates. \citet{Bollen2018} presented the Model Implied Instrumental Variables (MIIVs) approach which leverages the concept of instrumental variables (variables which lie upstream of causes and which are not directly related to the outcomes of interest) and two-stage least squares techniques. However, the MIIVs approach is less general than ours. This is because instrumental variables and two-stage least squares approaches assume linear functional form, whereas the ideas in this work depend solely on the conditional independence / Markov assumptions, which generalise to arbitrary and non-parametric dependencies. It is also worth noting that identifying valid instrumental variables can be challenging in itself \cite{Heckman1995, Angrist2001}, and does not necessarily provide a shortcut to clean results. However, here we certainly concede that the same challenges that apply to identifying and theoretically justifying instrumental variables also apply (and arguably more so) to the specification of any model, particularly as the complexity of the model increases.

In order to best communicate our approach, we begin with a brief review of the relevant background. As well as reviewing some generally relevant concepts, we aim to review four related concepts in particular, which will be combined in order to reduce the size of our structural models: conditional independence, Markov Blankets, projection, and identification. The application of these rules does not introduce any additional assumptions to those which already apply to the full model. Interested readers are encouraged to consult useful resources by \citet{Hunermund2021, Vowels2021DAGs, Cinelli2020, Peters2017, Koller2009, Kline2005, Pearl2009, Pearl2016}, and \citet{Loehlin2017}. In terms of notation, we use $X$ (or, \textit{e.g.} $A, B, C$ etc.) to denote a random variable, and bold font $\mathbf{X}$ (or, \textit{e.g.} $\mathbf{A}, \mathbf{B}, \mathbf{C}$ etc.) to denote a set of random variables. We use the symbols $\indep$ and $\not\!\perp\!\!\!\perp$ to denote statistical independence and statistical dependence, respectively. For linear systems, such statistical dependence may be identified using correlation, but the majority of our discussions are general and non-parametric. We use directed arrows to denote a directional structural/causal dependence, and $U$ (or $\mathbf{U}$) for a single (or set of) unobserved variable(s). Note that the theory we discuss is applicable to models with latent constructs (such as factor or measurement models), as well as those without (such as path and structural models), and generalises beyond linear models. This is because the theory we discuss is part of the general Structural Causal Modeling (SCM) and Directed Acyclic Graph (DAG) frameworks \citep{Pearl2009}. Path models and SEMs both represent a subset of the family of SCM and DAG models, where the \textit{functional} relationships between variables are assumed to be linear. In other words SCMs and DAGs make no assumptions\footnote{Strictly, there are some limitations here \citep{Maclaren2020}.} about whether one variable is an arbitrarily complex function of another. 

For example, in SCM terminology $A := f(B,C, U_A)$ indicates that A is some general function $f$ of $B$ and $C$. Here, $U_A$ tells us that $A$ is also a function of exogenous random variable $U_A$. Indeed, it is this $U_A$ which prevents the relationship between $A$ and  $B$ and $C$ from being deterministic. Structural Equation Models (SEMs), on the other hand, assume that all endogenous variables are the result of a linear weighted sum of others,  $A := \beta_{BA}B + \beta_{CA} C + U_A$. Here, the $\beta$s are structural parameters (also called path coefficients or effect sizes) which we wish to estimate. The walrus-shaped assignment operator $:=$ tells us that the left hand side is a structural outcome of the right hand side; the equations are not intended to be rearranged and there is very much a directional relationship involved.

\begin{figure}[t!]
\includegraphics[width=0.8\linewidth]{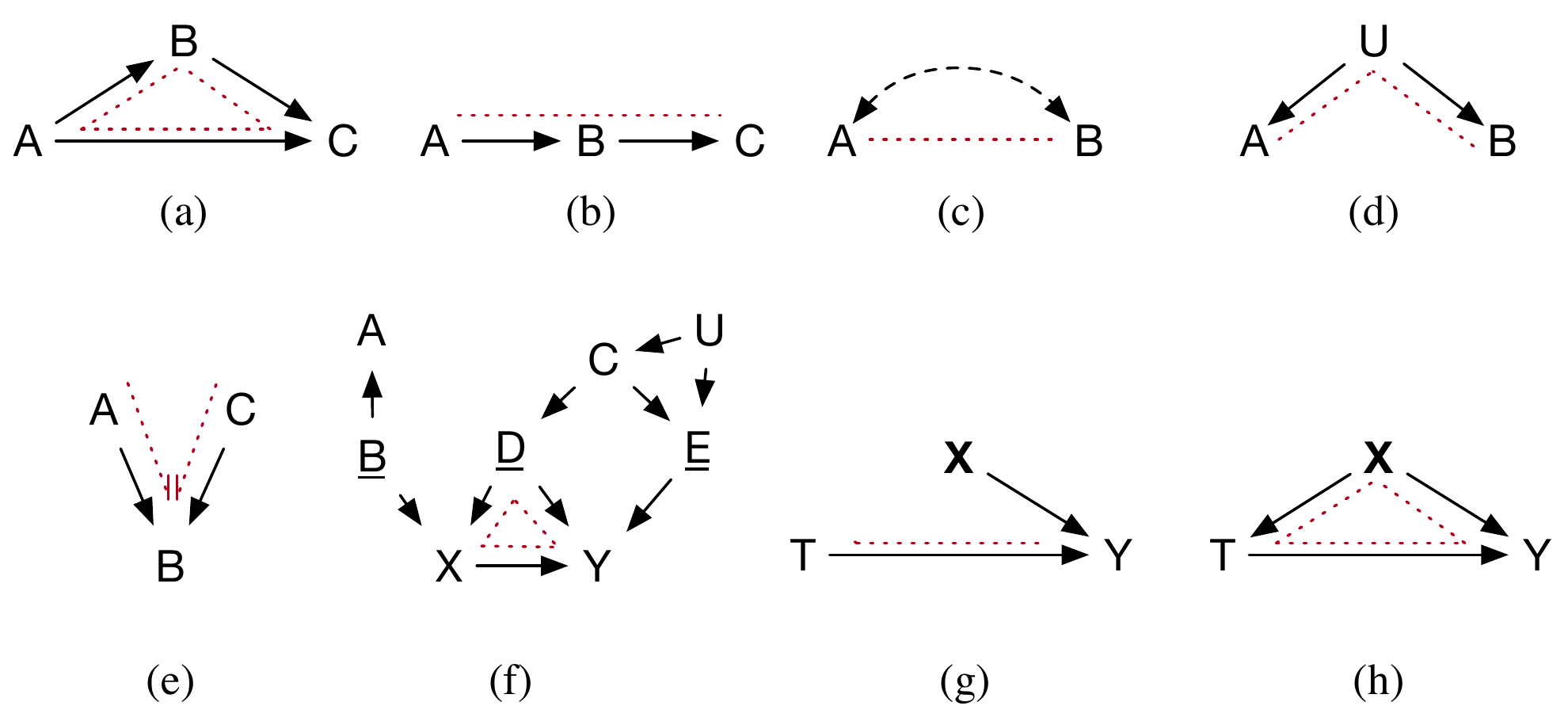}
\caption{This figure provide a number of example graphical models. Solid black lines indicate causal dependencies, dotted red lines indicate statistical dependence, parallel red bars indicate a `break' in statistical dependence (example (e)), \textbf{boldfont} indicates a set of variables, and the letter $U$ is reserved to denote unobserved variables.}
\label{fig:graphs}
\end{figure}

As we construct system of equations representing our SEM (or, indeed, our SCM) it is often convenient to represent these relationships graphically/visually. For example, consider the following set of (linear) structural equations:

\begin{equation}
    \begin{split}
        A := U_A,\\
        B := \beta_{AB} A + U_B,\\
        C := \beta_{AC} A + \beta_{BC} B + U_C.\\
    \end{split}
\end{equation}

These can be represented simply as the mediation model depicted in black, solid arrows in Figure~\ref{fig:graphs}(a). The variables $U$ are generally not included unless they are statistically dependent. Of course, this is frequently the case in psychology, and this may be denoted using a curved, bidirected edge, as between variables $A$ and $B$ in Figure~\ref{fig:graphs}(c), or by explicitly including the relationship as in Figure~\ref{fig:graphs}(d). Such relationships can, of course, also be included in the system of equations comprising the SEM. Note that, as a result of the causal structures present in the DGP, there are induced a number of statistical dependencies indicated in Figure~\ref{fig:graphs} by the red dotted lines.

\subsection{The Data Generating Process}
It is worth maintaining conceptual separation between: (1) the process occurring in the real world, which we might refer to as the Data Generating Process (DGP), (2) Our SEM, which we generally want to sufficiently match the process in the real world, and (3) the specification of a multiple linear regression. As we will come to see, it is the interaction between (1) and (2), and (1) and (3) which determines whether or not the parameters estimated according to (2) and (3) are meaningful and unbiased.

For example, if have a strong theory that the DGP is equivalent to a fully mediated process $A \rightarrow B \rightarrow C$, then we would be advised to employ a model which `agrees' with this structure. By agreement we mean that the model we use facilitates the unbiased estimation of the parameters of interest, in the event that our theory about the true process is correct. One option we have is to specify everything about our theory explicitly using an SEM. However, what we aim to show is that if we are primarily concerned with a subset of parameters then in some cases we can significantly reduce the complexity of our model without affecting the `agreement' of our model with the DGP. In the case of the full mediation, it is interesting to note, for example, that including a direct effect in the SEM (in addition to the indirect effect) does not bias our estimates of the indirect path parameters. This is actually an example of how \textit{increasing} the complexity of the SEM does not result in `disagreement' between the SEM and the real-world DGP. In this paper, however, we will primarily be concerned with \textit{reducing} the complexity.

\subsection{Conditional Independencies}
The visual graphs provide us with a way to directly read off the conditional independency structure of the model. Conditional independencies tell us whether the inclusion of additional information changes anything about our knowledge, and such conditional independencies hold if the graph is `Markovian' \citep{Peters2017}.\footnote{Note that in the presence of unobserved variables, the graph is said to be semi-Markovian.} Generally, we assume that SEMs and SCMs are Markovian, such that the conditional independencies in the graph hold in the data, and can be inferred directly from the structure of the graph itself. The ideas are crucial for what follows, because we will use conditional independencies to inform us about which variables we need to include or exclude in our SEM.

Starting with the example in the full mediation model of Figure~\ref{fig:graphs}(b), we see that variable $C$ cannot contain information about $A$ which is not shared through $B$. Therefore, if we already know $B$, knowing $A$ tells us nothing more about $C$ than we already knew. This renders $A$ statistically independent of $C$ given $B$, which can be expressed as: $A \indep C | B$. This is known as a conditional independence statement, because it tells us which sets of variables are independent of each other given a set of conditioning variables. It is worth noting that when we run a regression (logistic or otherwise) we are estimating some outcome \textit{conditioned on} some set of predictors. Running the regression $\mathbb{E}[C |B, A]$ from data generated according to a fully mediated DGP will result in the same consequences as above: the fact we have included $B$ means that the importance given to $A$ will be zero (notwithstanding finite sample deviations). Clearly, therefore, an understanding of the structure is therefore absolutely crucial for constructing the regression models \citep{VowelsOutrun}. For instance, if $A$ is a treatment variable and we do not recognise $B$ as a mediator, the inclusion of $B$ in the model will result in a negligible coefficient estimate for $A$ which may well mislead us to think the treatment is ineffective.

In order to generalise this result to other graph structures, it is worth committing some rules to memory. If a graph contains these substructures:

\begin{equation}
    \begin{split}
        A \rightarrow B \rightarrow C,\\
        A \leftarrow B \rightarrow C,
    \end{split}
    \label{eq:a}
\end{equation}

then knowing/conditioning on $B$ renders $A$ and $C$ statistically independent. Of course, without this conditioning, $A$, $B$, and $C$ are all statistically dependent. These two graphs are known, respectively, as a chain and a fork. One can start to write the complete list of conditional independencies which are implied by these two graphs:

\begin{equation}
    \begin{split}
        A \not\!\perp\!\!\!\perp B, \; \; A \not\!\perp\!\!\!\perp C, \; \; B \not\!\perp\!\!\!\perp C,\\
        C \indep A | B, \; \; C \not\!\perp\!\!\!\perp B | A, \; \; B \not\!\perp\!\!\!\perp C |A,\\
    \end{split}
\end{equation}

and so on. Alternatively, if a graph is structured as follows:

\begin{equation}
        A \rightarrow B \leftarrow C,
    \label{eq:b}
\end{equation}

we have what is known as a \textit{collider}. Unlike the examples in Eq.~\ref{eq:a}, variables $A$ and $C$ are actually already independent. A collider is also depicted in Figure~\ref{fig:graphs}(e), and the parallel vertical red lines depict the `break' in statistical dependence between $A$ and $C$. Furthermore, conditioning on $B$ in this structure actually \textit{induces} statistical dependence between $A$ and $C$ - a phenomenon known as explaining away \citep{Pearl2009, Pearl2016}. A corresponding list of conditional independency statements might, therefore, resemble:

\begin{equation}
    \begin{split}
        A \not\!\perp\!\!\!\perp B, \; \; B \not\!\perp\!\!\!\perp C,  \\
        A \indep C,\; \; A \not\!\perp\!\!\!\perp C | B,\\
    \end{split}
\end{equation}

Note that conditioning on \textit{descendants}\footnote{Variables are known as \textit{ancestors} of downstream \textit{descendants} if there exists a directed path between the variables. A direct descendent is also called a child, and the direct ascendant is called a parent.} of the variable $B$ in the two graphs depicted in Eq.~\ref{eq:a} can \textit{partially} render $A$ and $C$ independent (because it essentially contains critical information from $A$ via $B$). Similarly, conditioning on a descendent of the collider variable $B$ in Eq.~\ref{eq:b} can also render variables $A$ and $C$ \textit{partially} dependent. Of course, two variables are either dependent or not, and the partial terminology is used here to communicate that the effect of conditioning is not as strong as would be the case using $B$ itself, as opposed to one of its descendants.

\subsection{Markov Blanket}
The conditional independency rules introduced above can be used to define a Markov Blanket. Essentially, the blanket constitutes a set of variables which yield statistical separation between variables `within' the blanket,and those outside it. Consider Figure~\ref{fig:graphs}(f) which depicts a Markov blanket around variables $X$ and $Y$. The underlined variables $B$, $D$, and $E$ constitute the Markov blanket - knowing or conditioning on these variables renders $X$ and $Y$ independent of variables $A$ and $C$, which are outside of the blanket. Whilst not the case in this example, it is of course possible to have variables which fall in the Markov blanket which do not need to be conditioned on because of the presence of a collider structure already renders variables outside of the blanket defined by the collider as independent of those within the blanket.

An SEM model can be reduced in size to comprise only the variables and paths necessary in order to estimate set of paths of interest. Considering, again, Figure~\ref{fig:graphs}(f), if we are only interested in the path coefficients proximal to the variables $X$ and $Y$, we do not need variables $A$ or $C$, thus reducing the number of estimated paths from ten (if we include the paths from unobserved $U$) to five. We discuss more opportunities below.

\begin{figure}[t!]
\includegraphics[width=0.4\linewidth]{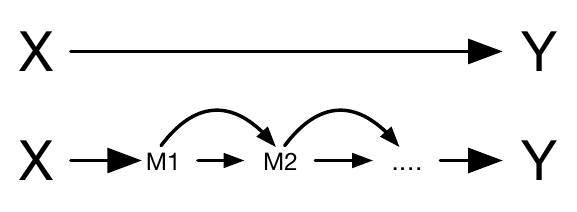}
\caption{This figure illustrates that between any two cause-effect pairs, there exists an almost infinitely decomposable chain of intermediate mediators.}
\label{fig:infmed}
\end{figure}

\subsection{Projection}
A cause-effect relationship can often be broken down into smaller and smaller subdivisions, until one starts talking about the effect of one molecule on the next in order to explain a simple game of billiards. As per Figure~\ref{fig:infmed}, each subdivision of the cause-effect relationships between $X$ and $Y$ could be represented as a mediating path with an infinite number of intermediate mediating paths. This makes it possible to ignore mediating variables (assuming there are no other structural implications), reducing, for example, $X \rightarrow M \rightarrow Y$ simply to $X \rightarrow Y$. Of course, such an extreme reduction may yield an intractably complex problem, and one might instead consider more modest examples, such as whether a treatment is mediated by some psychological mechanism(s). In this case, one can nonetheless reduce the problem (via projection) to an estimation of the total effect of treatment on the outcome, thus aggregating the intermediate direct and indirect effects and thereby reducing the complexity of the structure.

\subsection{Identification and Disentangling Statistical Influence}
Identification concerns whether or not, for a given graph, the causal effect we are interested in is actually estimable from the observed data \citep{huang2012pearls, Shpitser2008}. In the case where the full graph is given and there are no unobserved confounders, all causal effects are technically identifiable from the data. This means that there exists a mathematical expression which expresses the causal effect(s) of interest as a function of the observed statistical associations. If a causal effect is identifiable, it may be possible to estimate it with only a fraction of all the observed variables. Indeed, in such cases we may be more likely to use a multiple regression or indeed, some machine learning technique \citep{Vowels2020b, vanderLaan2011} to directly estimate the effect we care about.

Consider the graphs in Figure~\ref{fig:graphs}(g) and (h). Graph (g) represents the canonical Randomized Control Trial setup, where $T$ represents some treatment, $Y$ some outcome, and $\mathbf{X}$ some set of covariates which help to explain the outcome $Y$. In this graph, the covariates $\mathbf{X}$ are independent of treatment $T$ because of the random assignment of treatment. Such a structure means the only statistical dependence that exists between the treatment and the outcome is a result of the treatment itself. This statistical dependence is thus equivalent to the causal dependence we are interested in. As such, the effect can be directly estimated by comparing the outcome under different treatments.\footnote{Note that one may still wish to consider $\mathbf{X}$ too - it can be used to explain additional variance in $Y$ in order to tighten the estimate of the treatment effect.} 

In contrast, in observational studies patients may select their own treatment, and graph Figure~\ref{fig:graphs}(h) is more appropriate. For instance, if age is one of the covariates, older patients may prefer medication and have a lower chance of recovery, whilst younger patients may prefer surgery and have a higher chance of recovery. Thus, if we wish to estimate the \textit{causal} influence of treatment $T$ on the outcome $Y$, we need to somehow adjust for the additional statistical dependence that exists between $Y$ and $T$ which results from the `backdoor' non-causal path $T \leftarrow \mathbf{X} \rightarrow Y$. This is non-causal because there is no directed path between $T$ and $Y$ via $X$ (the arrow points from $X$ to $T$, not the other way around). Knowing the rules of conditional independencies described above, we know that for the substructure $T \leftarrow \mathbf{X} \rightarrow Y$, we can achieve $Y \indep T|\mathbf{X}$ in order to essentially simulate the structure of the graph Figure~\ref{fig:graphs}(h). In other words, by conditioning on $\mathbf{X}$ we `block the backdoor' path of \textit{confounding} statistical dependence which `flows' from treatment to outcome by conditioning on $\mathbf{X}$. This leaves only the one statistical path, which is also the causal path we care about. In this case, the statistical dependence is equivalent to the causal dependence we wish to estimate.

A detailed description of how to use identification is beyond the scope of this paper, but we will use some of these ideas to provide additional opportunities for SEM simplification.

% \subsection{Statistical Power}
% The sample size required for a given level of statistical power depends on the sizes of the effects we wish to estimate, the alpha level (\textit{e.g.}, 0.05) and the number of paths \citep{Loehlin2017}. In a large SEM with $K$ variables, the maximum number of paths is equal to $K(K-1)/2$ which, for $K=6$ means fifteen paths and fifteen corresponding path coefficients. Intuitively, the more paths we wish to estimate the more data are needed for a given level of statistical power, although the exact sample size also depends on the structure itself, and it is recommended to undertake simulations rather than use approximate rules of thumb \citep{Wolf2013}. As paths are removed, the number of degrees of freedom increases, and the less data are required. Hence, the goal of this paper follows quite naturally: removing paths reduces the required sample size for a given set of effects and a desired level of statistical power.

\section{Reducing SEMs - Worked Examples}

\begin{figure}[t!]
\includegraphics[width=0.9\linewidth]{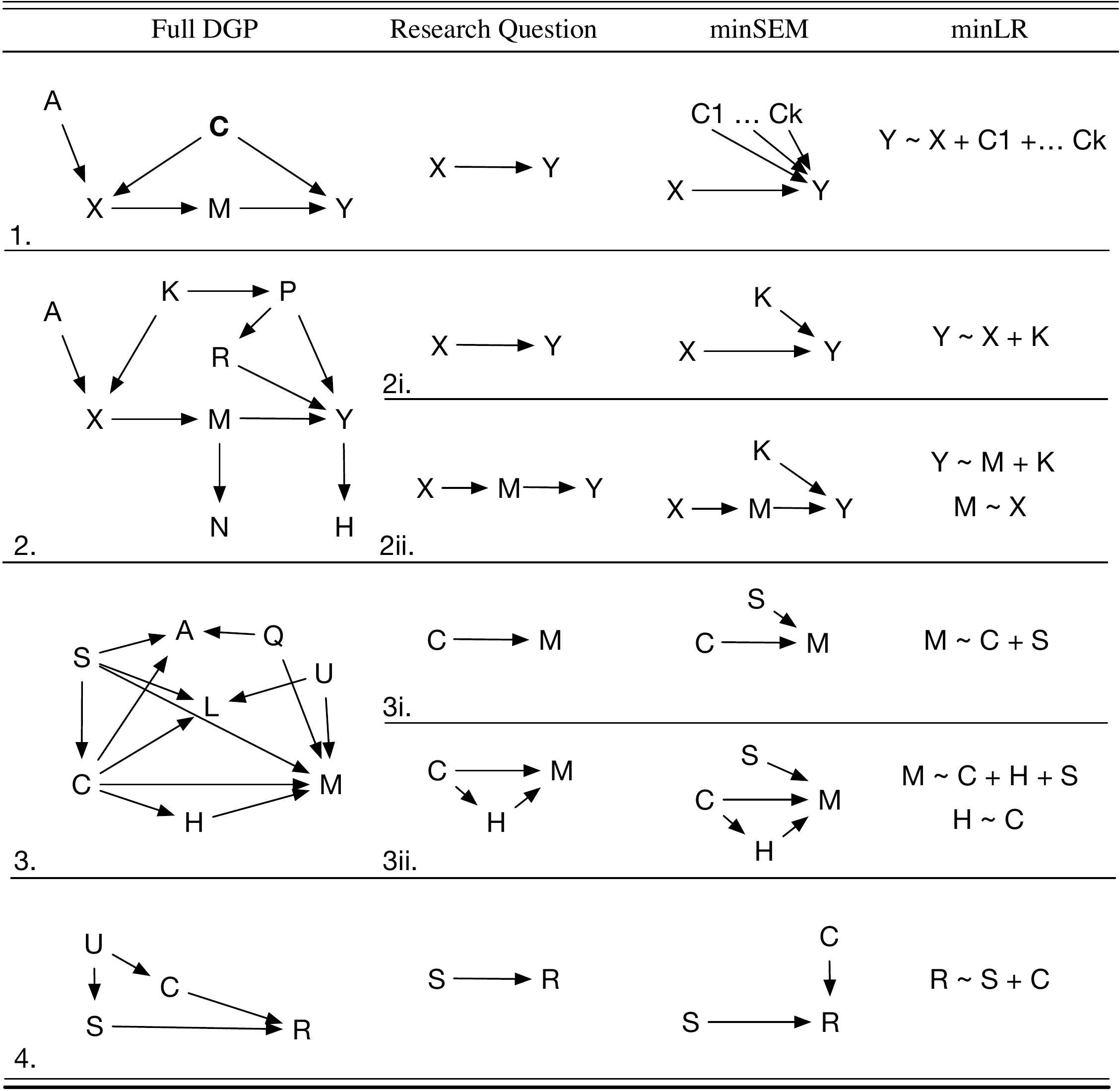}
\caption{ This figure presents a number of examples for taking the full `true' Data Generating Process (DGP) and finding the reduced graph and minimal linear/logistic regression (minLR) required to answer a given research question.}
\label{fig:all}
\end{figure}

In the previous section we reviewed four concepts which we will use for simplifying our SEMs without introducing bias into our effect estimates: (1) conditional independencies, (2) Markov Blankets, (3) projection, and (4), identification. In order to demonstrate these various techniques, we will walk through a number of examples which are presented in Figure~\ref{fig:all}. For each example, we specify (a) a full DGP as our starting point which we assume to be true and complete (`Full DGP' in Figure~\ref{fig:all}), (b) a set of paths of interest (`Research Question' in Figure~\ref{fig:all}), (c) a minimal SEM (denoted Reduced in Figure~\ref{fig:all}), and (d) syntax for the R \textit{lm()} function for a multiple regression. Five example DGPs are shown in Figure \ref{fig:all}.

In practice, our DGP will be developed using domain knowledge and/or causal discovery techniques \citep{Vowels2021DAGs, Vowels2021, Glymour2019}. For now, we provide general examples with a view to demonstrating the ways in which the concepts reviewed above can be used to reduce our SEM.  Similarly, in practice the set of paths of interest will be determined by our research questions and our hypotheses. Note that it may be possible to simplify SEMs using bearing in mind other techniques which are applicable to linear models (such as instrumental variables) \citep{Bollen2018}, but we focus on those techniques reviewed above because they are generally applicable to a much broader family of problems. Finally, it is worth remembering that if a set of variables and paths are not needed for the SEM, then we also do not need to collect these variables to begin with, thus saving additional time and expense which could be used to, for example, collect more samples of the variables that really matter.\footnote{Note that some variables may not strictly be necessary for the estimation of the effect but may nonetheless be worthwhile including. For example, proximal causes of an outcome which do not interfere with our estimation of other desired causes can increase the precision of our estimates, and may therefore still be worth including \citep{Cinelli2020}.}

 In order to motivate the examples, we will attempt to describe a semi-plausible DGPs for psychological processes, but note that these examples are primarily taken to illustrate the process of simplifying the SEM, and the specifics should therefore not be taken too literally. Let us discuss each of the examples in Figure~\ref{fig:all} in turn. Finally, we also provide simulation results for DGPs 2-5 in Figures \ref{fig:chi2}-\ref{fig:pvalmae}. We omit simulations for DGP 1 because it represents a reduction of the other examples, and so including it is somewhat redundant.

\subsection{Example 1: Mediated Treatment}

Starting with the first example depicted in Figure~\ref{fig:all}, let us begin by considering what this graph could possibly represent. Variable $Y$ could be an outcome (\textit{e.g.} depressive symptoms) for a therapy $X$, the effect of which is mediated by therapeutic alliance $M$. The set $\mathbf{C}$ represents covariates that influence the choice of therapy modality as well as the likelihood of recovery, and includes factors such as age, gender, history of mental health problems, and so on. Finally, variable $A$ could represent a personal attitude which influences the choice of treatment but which does not influence whether the person recovers.

For this example, let us assume that our research question concerns estimation of the efficacy of treatment on the outcome, \textit{i.e.}, $X\rightarrow Y$. The minimal SEM (denoted in Figure~\ref{fig:all} as minSEM) requires three fewer paths in order to estimate this question. Firstly, if we are not interested in the particulars of the mediated path $X \rightarrow M \rightarrow Y$ then we do not need to include $X\rightarrow M \rightarrow Y$, or to therefore collected data for $M$  (afforded by the projection concept reviewed above). Secondly, even though there exists a spurious/confounding/backdoor path $X \leftarrow \mathbf{C} \rightarrow Y$, we do not need to estimate the actual path $X \leftarrow 
\mathbf{C}$ so long as we include the path $C \rightarrow Y$. This is another application of the conditional independency rules. Thirdly, we do not need to include $A$ in the model (neither do we need to collect data for $A$). Adding the path $A \rightarrow X$ into the model is superfluous to the effect we are interested in. Finally, note that the resulting \textit{lm()} syntax contains only the two necessary variables as predictors - $X$ and $C$.

\subsection{Example 2: Structured Controls}

We use the term control variables to mean variables which we wish to adjust for and which would otherwise leave an opening for non-causal, statistical association. For example, the set of variables $\mathbf{X}$ in Figure~\ref{fig:graphs}(h) could be considered to be a set of relevant control variables which enables us to get unbiased estimation of the effect of treatment $T$ on the outcome $Y$. However, it is worth considering that a set of control variables itself may comprise a complicated structure in its own right, and we consider two cases, the second of which is discussed in the following section.

The first graph with structured controls is given as example 2 in Figure~\ref{fig:all}. We can consider the meaning of variables $A, X, M$, and $Y$ to be the same as in example 1, that is attitude, treatment, treatment-outcome mediator, and outcome, respectively. The difference now is that we also have a mediation child $N$, an outcome child $H$, and a structured set of control variables $K, P$, and $R$. If, as indicated in example 2i, we are only interested in estimating the effect of $X$ on $Y$ then, as in the first example, we can ignore $A$ and $M$, as in the previous example. Similarly, we can also exclude $N$ and $H$ for our minSEM, as their existence in the DGP does not change the principal relationship we are interested in. 

There still exists a backdoor path through the control variables $K, P, R$, and $Y$, and so we need to understand which of the associated variables and paths to include in our minSEM to adjust for this spurious path. There exist the following options which block this path: $K \rightarrow Y$, $K \rightarrow P \rightarrow Y$, and $R \rightarrow Y \leftarrow P$. Note that $R \rightarrow Y$ is not an option by itself because this would leave the path through $P\rightarrow Y$ open. Note also that we do not need to estimate the path $K \rightarrow X$ because we are not interested in this effect. Thus, overall, our minSEM reduces to the estimation of only two paths (reduced from ten), as in the previous example. The linear regression also remains equivalent.

If our research question involved the estimation of the mediation, as in example 2ii in Figure~\ref{fig:all}, then the only change to minSEM needs to be the inclusion of the mediation $X \rightarrow M \rightarrow Y$. The linear regression now involves two stages to decompose the problem into two sets of paths (one from $X\rightarrow M$, and the other comprising the paths $M\rightarrow Y$ and $K\rightarrow Y$).

\subsection{Example 3: Colliding Controls}
One might be forgiven for thinking that the safest thing to do with a set of control variables is to always include them in the model in order to make sure we are blocking the backdoor paths. In the previous example, for instance, we could just play it safe by including $\{K,P,R\}$. However, example 3 in Figure~\ref{fig:all} shows that some putative control variables may include collider structures. Let us consider that variables $C$, $M$, and $L$ are class-size, math exam score, and language exam score, respectively. $H$ represents a mediator such as whether a student does their homework, $S$ represents Social Economic Status (SES) - perhaps children with higher SES attend schools with smaller class sizes and have better grades overall - $U$ represents an unobserved attribute of intelligence $Q$ a measured attribute of intelligence, and $A$ musical ability. 

Based on example 3i we are interested in the effect of class size on math exam score. It might be tempting to include the paths concerning the other related scores (such as language score, or musical ability). In the case of musical ability, we \textit{could} include the paths $C \rightarrow A \leftarrow Q \rightarrow M$ without causing any problems, but it doesn't actually help us estimate the effect we are interested in. Indeed, the collider structure $C \rightarrow A \leftarrow Q$ prevents any backdoor information affecting our estimation of $C \rightarrow M$, so we do not need these paths in our model. Another collider exists between $C \rightarrow L \leftarrow U \rightarrow M$, and even though the structure is the same, the fact that $U$ is unobserved means we cannot and should not include $L$ in the model. Indeed, if $L$ were to be included (without $U$ as $U$ is unobserved) we would induce a spurious path linking $C$ to $M$ through $L$ and $U$. Thus, again, even though these might be tempting control variables which we might think would, at best, increase precision, and at worst do nothing, in fact they cannot be included owing to the collider structure with an unobserved variable.

Thus, we have no need to include paths relevant to $A$ or $L$ in our model. Including the path $Q \rightarrow M$ may improve the precision of our desired estimate, but it is not necessary. The partial mediation through $H$, if not part of our research question, does also not need to be included. The only path we have to be concerned about is $C \leftarrow S \rightarrow M$, and we can deal with the induced statistical path by simply including the path $S \rightarrow M$. In this case, the minSEM contains two paths, whilst the full model (including the unobserved paths) involves thirteen. The corresponding linear regression is equally simple, and only includes $C$ and $S$ as predictors.

If we are interested in the partial mediation of class size, homework, and math exam score, then we can simply augment the minSEM from example 3i to include this additional structure. The linear regression also changes to accommodate the estimation of the additional paths, as with example 2ii.

\subsection{Example 4: Simple Unobserved Confounding}

The final example is relatively straightforward. Here, $R, S$, and $C$ could represent relationship satisfaction, partner support, and communication style, respectively, where the unobserved confounder $U$ between support and communication. The unobserved confounder induces a non-causal statistical dependence between $S$ and $R$ through $C$, and the minSEM therefore needs to include the path $C \rightarrow R$. The linear regression, similarly, needs only $S$ and $C$ as predictors. 

 \begin{figure}[t!]
\includegraphics[width=0.9\linewidth]{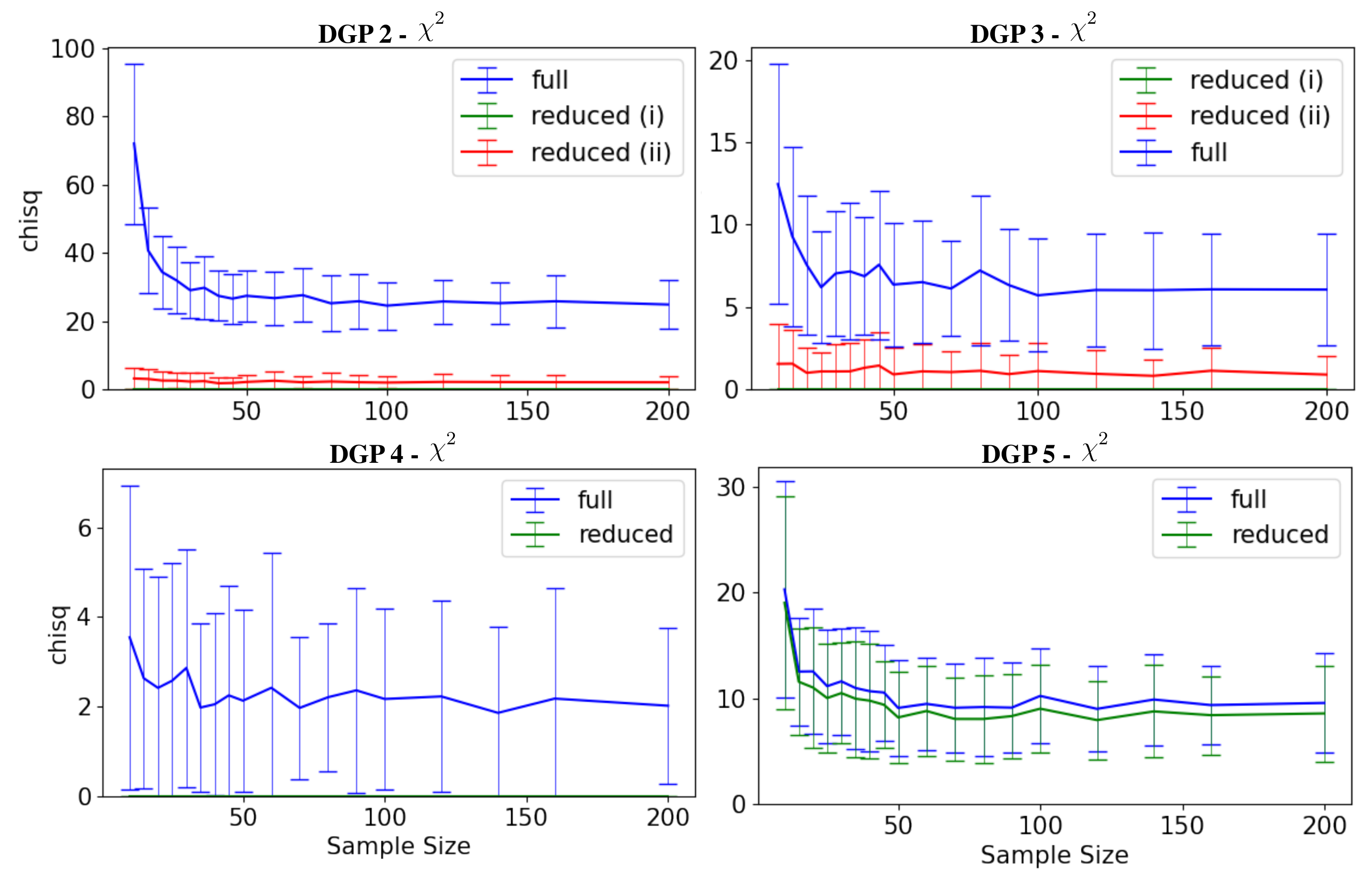}
\caption{ Averages and standard errors over 100 simulations with varying sample sizes for $\chi^2$ for data generated from Data Generating Processes (DGPs) 2-5 in Figure \ref{fig:all}.}
\label{fig:chi2}
\end{figure}

 \begin{figure}[t!]
\includegraphics[width=0.9\linewidth]{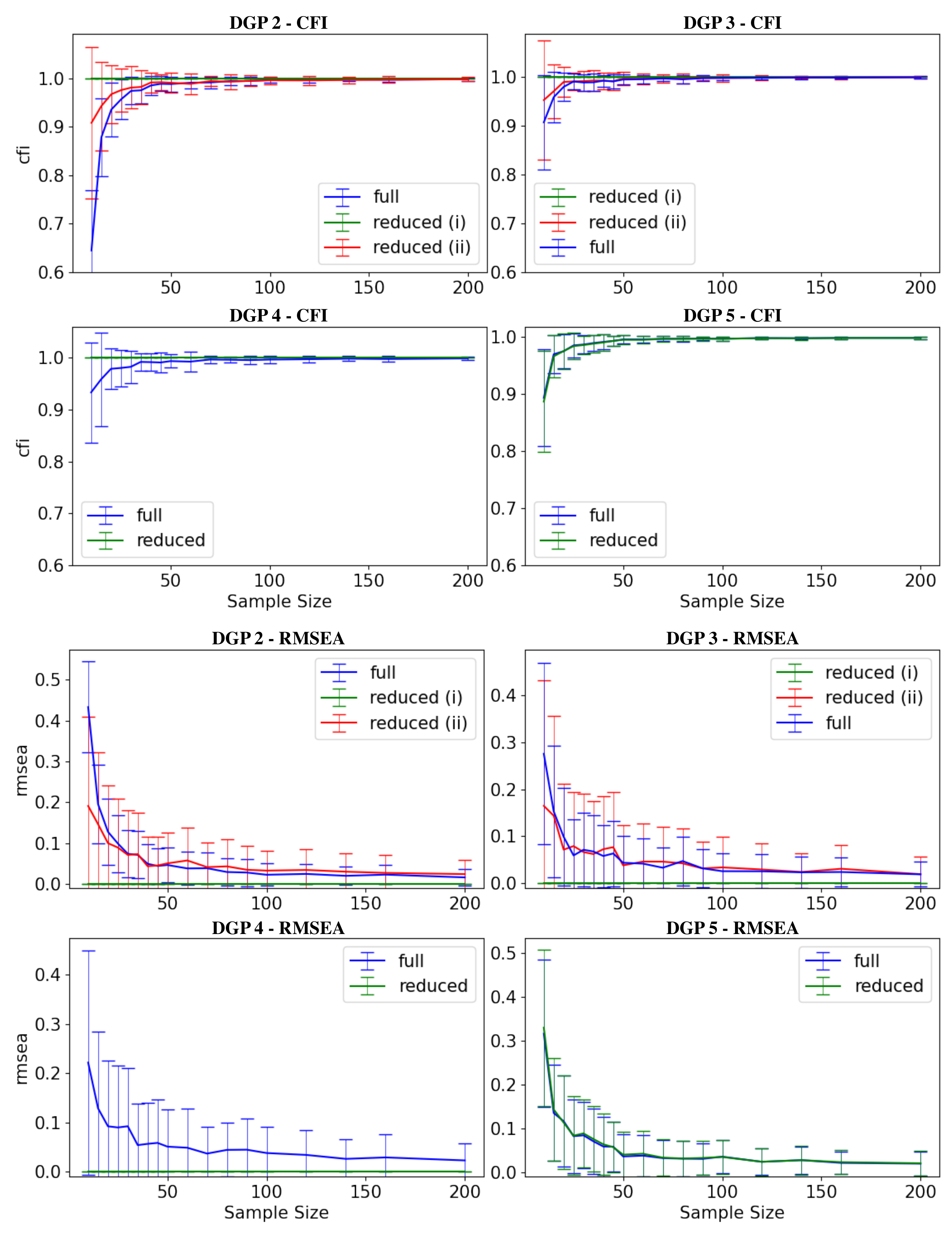}
\caption{Averages and standard errors over 100 simulations with varying sample sizes for Comparative Fit Index (CFI) and Root Mean Squared Error of Approximation (RMSEA) for data generated from Data Generating Processes (DGPs) 2-5 in Figure \ref{fig:all}.}
\label{fig:cfirmsea}
\end{figure}

 \begin{figure}[t!]
\includegraphics[width=0.9\linewidth]{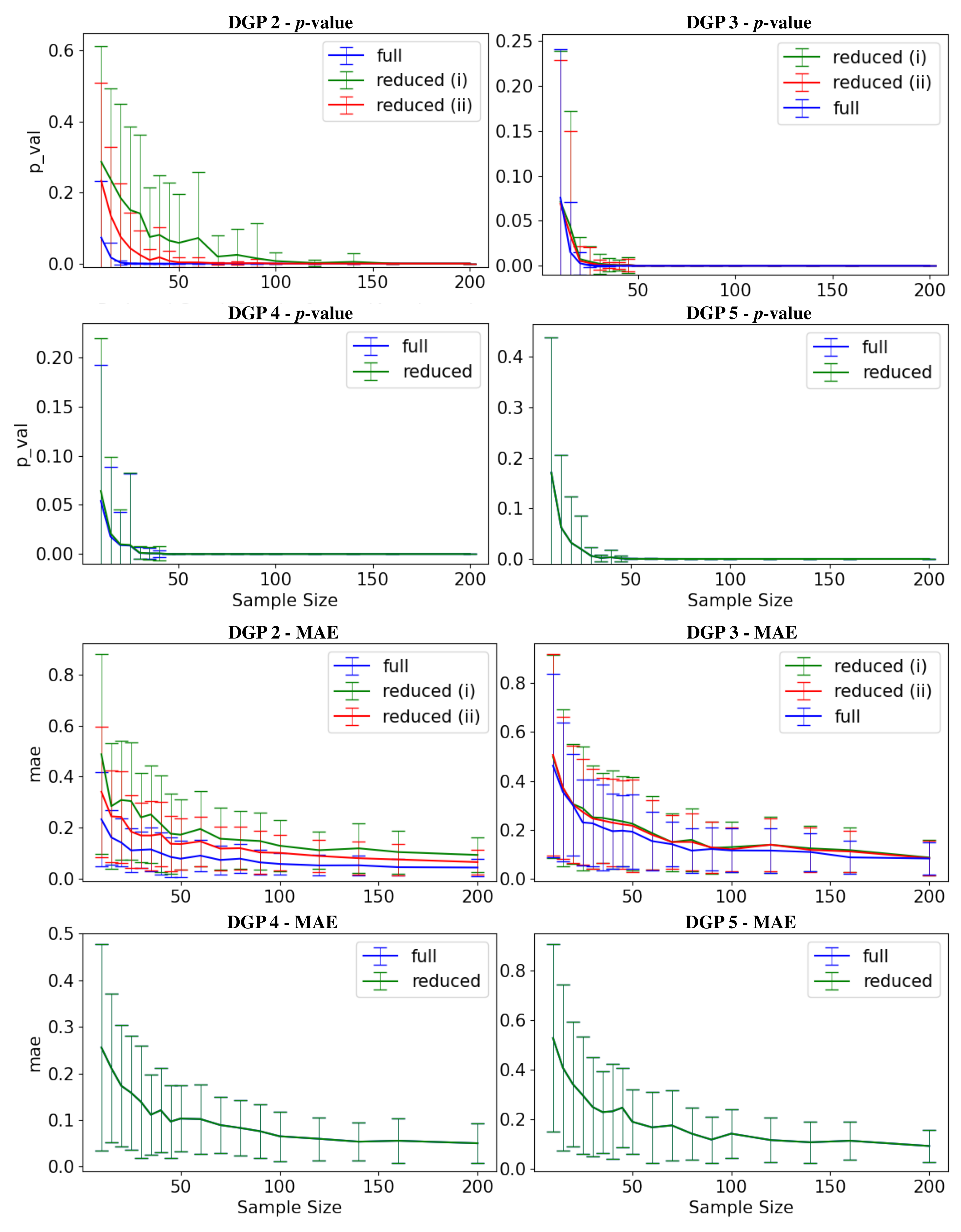}
\caption{Averages and standard errors over 100 simulations with varying sample sizes for \textit{p}-values and Mean Absolute Error (MAE) for data generated from Data Generating Processes (DGPs) 2-5 in Figure \ref{fig:all}.}
\label{fig:pvalmae}
\end{figure}

\subsection{Simulation Results}
Simulation results for DGP examples 2-5 in Figure \ref{fig:all} are shown in Figures \ref{fig:chi2}-\ref{fig:pvalmae}. We use the sem function in the lavaan library \citep{Rosseel2012} to estimate a single target effect for each variant.  We provide results for the following metrics: $\chi^2$, Root Mean Squared Error of Approximation (RMSEA), Comparative Fit Index (CFI), Mean Absolute Error (MAE) and \textit{p}-values. For the latter two, we provide results for a single effect of interest. For example, for the DGP research question 2ii in Figure \ref{fig:all}, we specify the SEM models given in the `Full DGP' and `Reduced' columns and generate MAEs and $p$-values for the total effect of $X$ on $Y$. Similarly, for DGP research question 3ii, we specify the SEM models given in the `Full DGP' and `Reduced' columns, and generate MAEs and $p$-values for the total effect of $C$ on $M$. Finally, for example 5, we specify the SEM models given in the `Full DGP' and `Reduced' columns, and generate MAEs and $p$-values for the total effect of $A1$ on $B3$. 

For each of the example DGPs, we generate data across a range of sample sizes (10-200), and for each sample size we undertake 100 simulations. The results of these 100 simulations are used to derive means and standard deviations for each of the metrics, thus allow us to compare the results when specifying the full DGP model compared with the reduced models.

Starting with the results for the model fit metrics $\chi^2$ in Figure \ref{fig:chi2}, we see that for DGPs 2-4 the reduced models all perform better (lower $\chi^2$ indicates better fit). Note that $\chi^2$ is a measure of model fit which accounts for the complexity of the model \citep{Loehlin2017, Kline2005}. It is therefore little surprise that the reduced models perform better in this regard, and it is also not a surprise that the differences for the full and reduced models for DGP 5 were not different - the reduced model did not differ greatly in its reduction of complexity. This is the first empirical confirmation that reducing the complexity of the model does not adversely affect the adjusted fit metric $\chi^2$.

Similar results are found for CFI and RMSEA in Figure \ref{fig:chi2}. The results indicate tat the reduced models achieve, at worst, equivalent performance, and at best, improved performance. The RMSEA shows greater improvement with the reduced model although, as with $\chi^2$, RMSEA is an adjusted metric, and so this is consistent with the expectation of a lower RMSE for the smaller models.

Finally, the \textit{p}-values and MAEs for the target effect size estimates are shown in Figure \ref{fig:pvalmae}. For DGP 2 (top left plot), the $p$-values are higher for the reduced model than the complete model. This is consistent with the expectation that the inclusion of more variables can help increase the precision of our estimates. Specifically, the reduced model in DGP 2i only includes two effects of the outcome $Y$, which is $X$ and $K$. However, other more proximal variables $P$ and $R$ exist, and their inclusion would improve the quality of the estimate. In this case, $R$ and $P$ would be doubling as both control variables (adjusting for the backdoor path from $X$ to $Y$, as well as variables which aid in precision \citep{Cinelli2020}. Note also that the standard deviation of these \textit{p}-values is higher, indicating greater variation across simulations. This increased variance also results in a higher MAE, which is also evidence in the DGP2 - MAE plot in Figure \ref{fig:pvalmae} (third row, first column). Thus, even though the effect size estimates will be unbiased (owing to correct specification of the reduced model with respect to the full DGP), the removal of explanatory variables can impact the precision of the estimates. In order to compensate for this, one can choose to retain variables which have explanatory power so long as their inclusion does not contradict the full, underling model. DGP 2 represents a useful example insofar as variables $R$ and $P$ can be included (optionally in addition to $K$), to help explain the effect of $X$ on $Y$.

\section{Summary and Conclusion}

We have provided a number of didactic examples showing that if we are presented with a specific question regarding a relatively complex process, we can simplify our SEMs considerably. The simplification process takes advantage of a number of graphical rules, and does not introduce any additional assumptions to those which already apply to the full model. In one of the demonstrative examples example, an SEM with upwards of thirteen paths was reduced to only two. The simulation results highlighted unsurprising improvements in adjusted model fit metrics (unsurprising because simpler models are penalised less than complex models according to such metrics). 

In terms of limitations, researchers should be mindful that the success of the approach rests on the degree of correct specification achieved when the DGP model is constructed. However, this limitation applies to \textit{all} statistical approaches which concern the estimation of interpretable / causal effects, and this approach does not alleviate the consequences of model misspecification. Furthermore, reducing model complexity may reduce the precision of the estimation because less explanatory power may be available to estimate an effect. This downside is somewhat offset by the possibility that, with a simpler model, a larger sample size may be acquired for equivalent cost. For example, if the simplification process indicates that a number of constructs with large inventories are no longer required, we may gain back significant data collection time which can be put towards the recruitment of more participants. Such possibilities therefore enable us to increase statistical power for estimating the effects we really care about. 

Even without the simplification process, translating a psychological theory into a graph is a worthy exercise, particularly when undertaken \textit{before} the data collection stage. It helps us be transparent and unambiguous about our model and assumptions, increases specificity for preregistration, and can highlight potential methodological challenges and difficulties before any resources have been expended. It may even highlight cases where estimation is not possible, and this relates to the problem of identification reviewed in the previous section. For example, if there exists an unobserved confounder between $X$ and $Y$ in the graph $X \rightarrow Y$, \textit{i.e.} $X \leftarrow U \rightarrow Y$, the causal effect cannot be estimated because the non-causal statistical association induced by the confounder cannot be adjusted for without access to $U$. These problems can, again, be seen by an inspection of the graph, and it is worthwhile identifying these problems sooner rather than later. In practice, such problems may be common, and either a researcher must do all they can to account for the possible unobserved confounders, or they must assume that a sufficient number have been already collected to assume that the problem is `ignorable' \citep{Pearl2009}. In general, it is important to remember that the goal of estimating causal effects rests on a number of strong (and often untestable) assumptions. However, it is only by taking causality seriously that we can understand what these assumptions are and whether they are reasonable.

In summary, SEM provides a powerful framework with which to encode our domain knowledge about a particular phenomenon of interest. In this paper we showed that, by using the concepts of conditional independencies and Markovicity, we can significantly shrink the required model without affecting the validity of the associated estimates, thereby reducing the required sample size and enabling us to redirect resources and funds towards the collection of data for variables which are critical to answering the questions we care about.

\vskip 0.2in
\bibliography{NN}

\end{document}